\documentclass[12pt,preprint]{aastex62}

\usepackage{gensymb}

\newcommand{\beq}{\begin{equation}}
\newcommand{\eeq}{\end{equation}}
\newcommand{\bea}{\begin{eqnarray}}
\newcommand{\eea}{\end{eqnarray}}
\newcommand{\p}{\partial}
\newcommand{\mb}{\mathbf}

\shorttitle{Particle acceleration in regions of magnetic flux emergence}
\shortauthors{Isliker, Archontis and Vlahos}

\begin{document}


\title{PARTICLE ACCELERATION AND HEATING IN REGIONS OF MAGNETIC FLUX EMERGENCE 
}


\author{H. Isliker}
\affil{Section of Astrophysics, Astronomy and Mechanics, Physics Department, Aristotle University,\\
    Thessaloniki, GR 541 24, Greece}
\email{isliker@astro.auth.gr}

\author{V. Archontis}
\affil{School of Mathematics and Statistics, University of St Andrews, \\North Haugh, St Andrews, Fife KY16 9SS, UK}
\email{vasilis@mcs.st-and.ac.uk}
\author{L. Vlahos}
\affil{Section of Astrophysics, Astronomy and Mechanics, Physics Department, Aristotle University,\\
    Thessaloniki, GR 541 24, Greece}
\email{vlahos@astro.auth.gr}





\begin{abstract}
The interaction between emerging and pre-existing magnetic fields in the solar atmosphere can trigger 
several dynamic phenomena, such as eruptions and jets. A key element during this interaction is the formation of large scale current sheets and, eventually, their fragmentation that leads to the creation of a strongly turbulent environment. In this paper, we study the kinetic aspects of the interaction (reconnection) between emerging and ambient magnetic fields. We show that the statistical properties of the spontaneously 
fragmented and fractal electric fields are responsible for the efficient heating and acceleration of charged particles, which 
form 
a power law tail at high energies on sub-second time scales. A fraction of the energized particles escapes from the acceleration volume, with a super-hot component with temperature close to $150\,$MK, and with a power law high energy tail with index between -2 and -3. We estimate the transport coefficients in energy space from the dynamics of the charged particles inside the fragmented and fractal electric fields, and the solution of a fractional transport equation, as appropriate for a strongly turbulent plasma, agrees with the test particle simulations. 
We also show that the acceleration mechanism is not related to Fermi acceleration,
and the Fokker Planck equation is inconsistent and not adequate as a transport model. 
Finally, we address the problem of correlations between spatial transport and transport in energy space.
Our results confirm the observations reported for high energy particles (hard X-rays, type III bursts and solar energetic particles) during the emission of solar jets. \end{abstract}


\section{Introduction}
Emerging magnetic flux is one of the mechanisms responsible for the formation of large scale reconnecting current sheets in the solar corona. The evolution of the reconnecting current sheet is proposed as the key mechanism for many explosive phenomena in the solar atmosphere, i.e.\ flares, prominence eruptions, jets, coronal mass ejections \citep{Heyvaerts77, Archontis04, Galsgaard05,Archontis04, Archontis05, Archontis12a, Archontis12b, Archontis13,Moreno-Insertis13,Karibabadi2013c,Jiang16,  Raouafi16, Wyper16, Wyper17}. Most of the cited studies focus on the large scale formation of the current sheet, the complex magnetic environment around the unstable current structure, the ejected plasma, and the jets in the vicinity of the current sheet.  

Several RHESSI observations of the base of coronal jets are associated with Hard-X Ray (HXR) emission   \citep{ Bain09, Glesener12, Glesener2018}. Frequently during coronal jets, the temporal profile of the associated HXRs matches the associated type III radio bursts \citep{Chen13}. Impulsive solar energetic particle events are also related to the jets (see the review by \cite{Raouafi16}). It is then obvious that jets act as an efficient mechanism for the heating and acceleration of particles, mainly due to the reconnecting current sheets at the boundary between the emerging magnetic flux and the ambient magnetic field in the solar atmosphere.

In a series of separate studies the  evolution of the unstable current sheets has been analysed in detail using Magnetohydrodynamic (MHD) and Particle In Cell (PIC) simulations.  The main common observation is that the reconnecting current sheet will fragment, forming a very efficient particle acceleration environment \citep{Onofri06,  Drake06, Kowal11, Hoshino12b, Cargill12,Lazarian12, Baumann13, Karibabadi2013c,  Guo15}. The results of these studies were limited to the analysis of the evolution of particles inside very small scale PIC simulations with periodic boundary conditions, or test particle simulations in the fields generated through the solution of the resistive MHD equations,  or by using the Fokker-Planck equation with analytically estimated transport coefficients \citep{Drake13, Guo15} 
	or  the comprehensive Fokker-Planck transport equations 
	that were developed recently
	for studying energetic particle 
	transport and acceleration in plasma regions containing numerous dynamic
	small-scale flux ropes 
	 \citep[e.g.][]{Zank2014,leRoux2015,leRoux2018}. 
Several recent studies based their analysis of the acceleration of particles in fragmented current sheets on the findings of PIC simulations or on simplified analytical models for the interaction of the particles with magnetic blobs resulting from the evolution of the current sheets. They assume that the fragments of the evolving current sheet are uniformly distributed in space, and the main acceleration mechanism is first order Fermi acceleration in a periodic simulation box \citep{Kowal11, Lazarian15}, 
and in particular they consider acceleration by island contraction, either in a compressible \citep[][which indeed leads to a first order Fermi process]{Zank2014, leRoux2015, Li2018} or an area-preserving way \citep[][which actually results in a second order Fermi process]{Dahlin16}. 
All these assumptions are open for discussion when one considers 
the more realistic fragmentation of the large scale current sheets 
in the solar corona, which are three dimensional open systems,
and when one relies on the dynamic evolution of the particle orbits in order to test the validity of the FP equation, as we do it in our approach. 

In the present paper, we focus on the kinetic aspects of the interaction between an emerging and a pre-existing magnetic field in the solar corona. For this study, we use an MHD numerical simulation, similar to the work by \cite{Archontis12a,Archontis13}, which shows: (i) the formation and fragmentation of large-scale current sheets during the emergence and interaction phase, and (ii) the emission of ``standard'' and more explosive eruption-driven ``blow out'' jets \citep{Moore10}. 
Our analysis explores the statistical properties of the electric field in the fragmented reconnecting current sheets, and the resulting energy distribution of electrons, by using a test particle numerical code. In our analysis, the particle dynamics inside the fragmented electric fields is directly used to estimate the transport coefficients of the particles, and we solve the transport equation that is appropriate for such an environment.

We address three important questions in this article:
\begin{itemize}
\item What are the statistical properties of the electric fields associated with the fragmentation of the large scale reconnecting current sheet~?
\item What are the characteristics of the electron energy distribution driven by the fragmented electric fields~?
\item Are the transport properties of the electrons inside the fragmented current sheet ``normal'' or ``anomalous''~? (We consider both, energy- and position-space.)
\end{itemize}

In section 2, we briefly present the MHD model used in this study and the test particle code. In section 3, we analyze the statistical properties of the fragmented electric fields. In section 4, we use the test particle code to follow the evolution of the energy distribution in the vicinity of the standard jet, and in section 5 the one near the base of the blow out jet. In section 6 and 7 we examine the transport properties of the particles in energy- and in position-space, respectively, and in section 8 we summarize our results.

\section{The model}

\subsection{The MHD model}
\begin{figure}[!ht]
\includegraphics[width=0.9\textwidth]{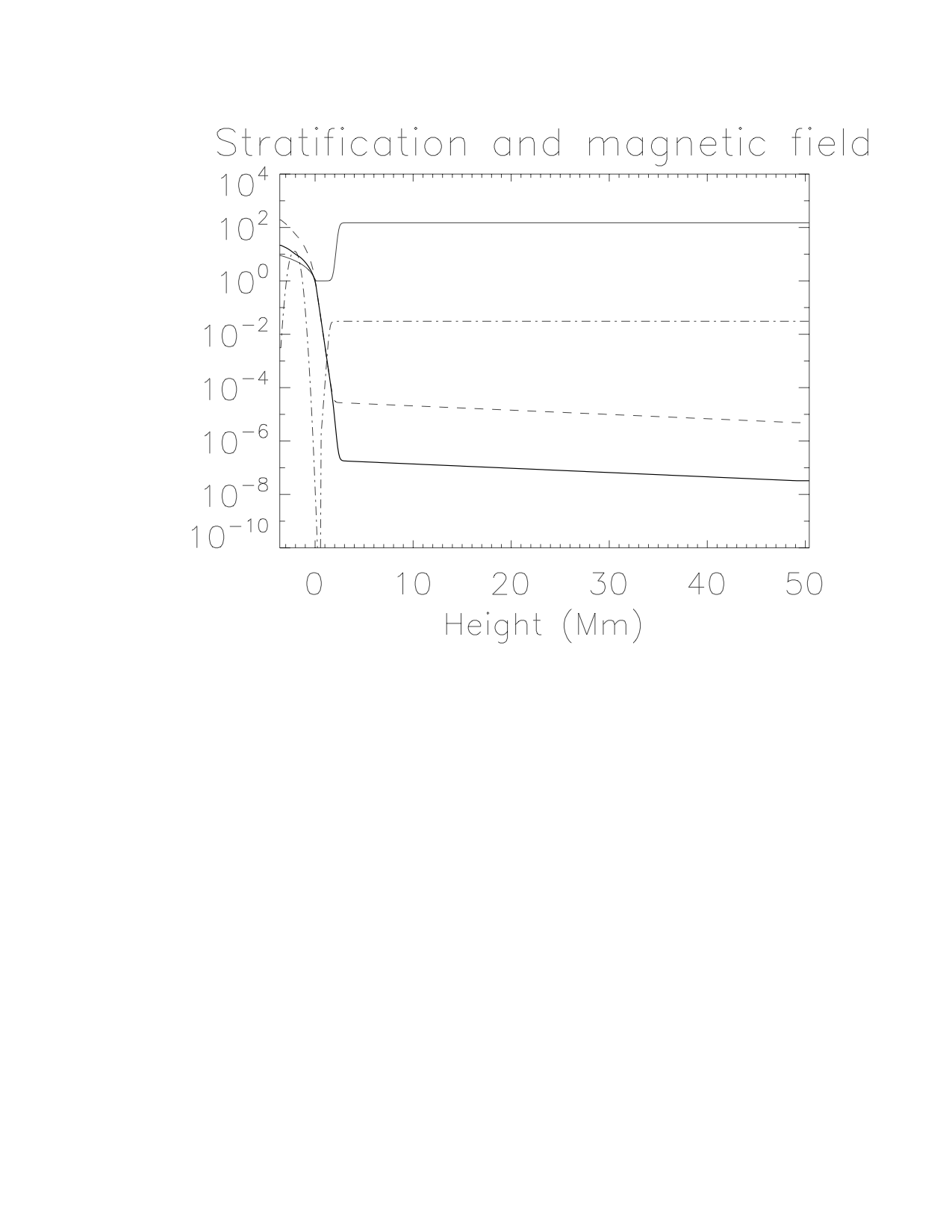}
\caption{The initial stratification of the atmosphere and the magnetic field 
in dimensionless units: temperature (thin solid), gas pressure (dashed), density (thick solid), magnetic field (dot-dashed). The dimension units we use are temperature $T=5100~\mathrm{K}$, pressure $P=7.16\times 10^3\ \mathrm{erg}\ \mathrm{cm}^{-3}$, density $\rho=1.67 \times 10^{-7}\ \mathrm{g}\ \mathrm{cm}^{-3}$, and magnetic field strength $B=300 \ \mathrm{G}$.}
\label{stratif}
\end{figure}

For this model, we are using the Lare3d code \citep[]{Arber01}. We solve the 3D time-dependent, resistive \& compressible MHD equations in
Cartesian geometry, as in the model by \cite[]{Archontis13b}.
Initially, the plasma is embedded into a plane-parallel hydrostatic atmosphere.
A highly stratified atmosphere in hydrostatic equilibrium is included in the model (see Fig.\ \ref{stratif}). The atmosphere consists of various layers. The solar interior is modeled by an adiabatically stratified layer, which resides in the range ($-3.6\,\mathrm{Mm} \leq z < 0\,\mathrm{Mm}$). Above it, the layer is isothermal ($5100\,\mathrm{K}$) and then the temperature increases smoothly with height up to $\approx 3\times10^{4}\,\mathrm{K}$. This layer represents the photosphere/chromosphere and it is located at $0\,\mathrm{Mm}\leq z < 1.9\,\mathrm{Mm}$. Then, at $1.9\,\mathrm{Mm}\leq z \leq 3\,\mathrm{Mm}$, the temperature increases with height, forming a layer that represents the transition region. The top layer in the stratified atmosphere, is an isothermal layer  ($\mathcal{O}(1)\,\mathrm{MK}$)  at $3\,\mathrm{Mm} < z \leq 50.4\,\mathrm{Mm}$, which is mimicking the solar corona. 
In the solar interior, at $z_{0}=-2.1\,\mathrm{Mm}$, we have included a twisted magnetic flux tube oriented along the $Y$-axis. The tube's magnetic field is defined by
\begin{equation}
B_y=B_0\,\mathrm{exp}(-r^2/R^2), \,\,\,\,\, B_\theta=\alpha\,r\,B_y,
\end{equation}
where $B_{y}$ is the longitudinal component of the magnetic field (i.e. along the axis of the tube) and $B_{\theta}$ is the azimuthal component. $B_{0}=3.9\,\mathrm{kG}$ is the initial field strength of the tube and   $\alpha=2.2 \times 10^{-3}\,\mathrm{km^{-1}}$ is the parameter associated with the uniform twist around the axis of the tube. With this twist, the tube is marginally stable to the kink instabilty. The radius of the tube is $R=450\,\mathrm{km}$ and $r$ is the radial distance from the axis of the tube
($r^{2} = x^{2} + (z + z_{0})^{2}$). To initiate the emergence of the tube, we impose a density along the axis of the tube so that its central part is more buoyant than its footpoints:
\begin{equation}
\Delta\,\rho=[p_t(r)/p(z)]\,\rho(z)\,\mathrm{exp}\,(-y^2/\lambda^2),
\end{equation}
where $p_t$ is the pressure within the tube and $\lambda$ defines the length of the buoyant part of the emerging field. In this model, we use $\lambda=0.9\,\mathrm{Mm}$.

The atmosphere in this model is magnetized. More precisely, there is an ambient magnetic field in the corona, with an oblique orientation, and it is defined by
\begin{equation}
B_c=B_c(z)\,(0,\cos\theta,\sin\theta),
\end{equation}
where $\theta=80\degree$ is the angle that the ambient field makes with respect to the positive $y$-axis (i.e.\ a vertical field has $\theta=90\degree$). The relative orientation between the emerging field and the ambient field is such that effective reconnection occurs when the two fields come into contact. The strength of the ambient field is $B_c(z)=9\,\mathrm{G}$ above the photosphere, and it gradually decreases to 0 
under the solar surface.

The numerical domain is $[-23.4,23.4] \times [-25.2,25.2] \times[-3.6,50.4]\,\mathrm{Mm}$ in the transverse ($x$), longitudinal
($y$) and vertical ($z$) directions, respectively. The numerical grid has 384 nodes in the transverse ($x$) direction, 416 nodes in the $y$ direction and 512 nodes along the height. Periodic boundary conditions have been implemented along the transverse and longitudinal directions, and open boundary conditions at the top of the numerical domain. The bottom boundary is a non-penetrating, perfectly conducting wall.



\subsection{The kinetic model\label{kinmod}}

The kinetic aspects of the dynamics of particles in an environment of fragmented current sheets is explored here by performing test-particle simulations in the electromagnetic fields of the MHD simulations.
The relativistic guiding-center equations of motions are used in the test-particle simulations, 
as given in
\cite{Tao2007}, which re-derive the relations of
\cite{Grebogi1984}, and we re-order the expressions to bring 
the equations into the form of \cite{Hamamatsu2007},
\beq
\frac{d\mb{r}}{dt}= \frac{1}{B_{||}^*} 
\left[ \frac{u_{||}}{\gamma} \mb{B}^* + \hat{\mb{b}}\times 
\left(\frac{\mu}{q\gamma}\nabla B -\mb{E}^* \right) \right]
\eeq
\beq
\frac{du_{||}}{dt} = - \frac{q}{m_0 B_{||}^*}\mb{B}^* 
\cdot\left(\frac{\mu}{q\gamma} \nabla B -\mb{E}^* \right)
\eeq
with $\mb{r}$ the particle position, and where $\mb{B}^*$ and $\mb{E}^*$ are the modified fields, defined by
\beq
\mb{B}^*=\mb{B} +\frac{m_0}{q} u_{||}\nabla\times\hat{\mb{b}}
\eeq
\beq
\mb{E}^*=\mb{E} -\frac{m_0}{q} u_{||} \frac{\p\hat{\mb{b}}}{\p t}
\eeq
with $m_0$ and $q$ the particle's rest-mass and charge, respectively,
$\mu$ the magnetic moment  
\beq
\mu = \frac{m_0 u_\perp^2}{2 B}  
\eeq
and $\gamma$ the Lorentz factor
\beq
\gamma = \frac{1}{\sqrt{1-(v/c)^2}}
=\sqrt{1+\frac{u^2}{c^2}}  .
\eeq
The evolving velocity variable $u_{||}$ is the parallel component  of the 
four-velocity $\mb{u}=\gamma\mb{v}$ ($\mb{v} = d\mb{r}/dt$), and $u_\perp$ is  the perpendicular component of the latter.

The electromagnetic fields of the MHD simulations are 
interpolated  in three dimensions with local third order polynomials (tri-cubic interpolation), which are continuous 
over the grid-points in the components and the derivatives up to first and partially to second order.
The parallel electric field is interpolated explicitly,
i.e.\ it is not calculated in-between the grid-points from the 
interpolated three Cartesian electric field components, since with this procedure we  avoid effects of artificial particle energization.
The temporal integration of the equations of motion is done with the 
Dormand-Prince scheme of the family of Runge Kutta methods, with adaptive time-step.

Collisions, when taken into account, are implemented as a Monte Carlo method, 
i.e.\ we consider the collisions as a 
stochastic process and superpose it to the deterministic motion,
as out-lined by \cite{Hamamatsu2007} (see also \cite{Karney1986}).


\section{Statistical properties of the electric field activity}

\subsection{Overview over the MHD snapshots}

\begin{figure}[!ht]
\includegraphics[width=0.5\textwidth]{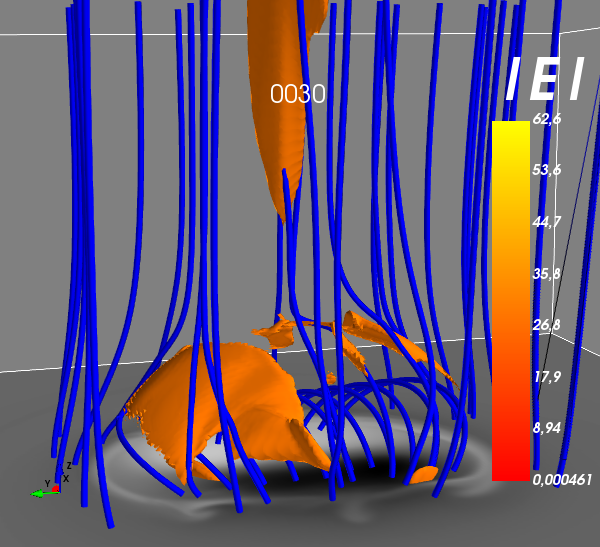}
\includegraphics[width=0.5\textwidth]{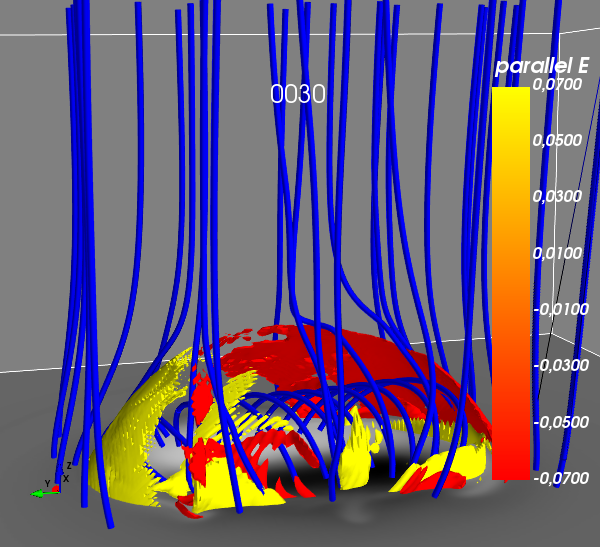}
\caption{
Results from the MHD simulations, closeup of the coronal part, snapshot 30. 
The left panel shows a visualization of selected magnetic field lines (blue) 
together with an iso-contour plot 
of the total electric 
field (orange 3D iso-surfaces). The vertically oriented isosurface (orange) is aligned with the direction of the reconnected fieldlines and it indicates the emission of the standard jet.
The horizontal $x$-$y$-plane shows  
the photo-spheric component $B_z$ as a 2D filled contour plot. 
The electric field is in physical units [V/m].
In the right panel, iso-contours of the parallel electric field are shown, indicating the fragmentation of the current sheet(s) at the interface between the interacting fields.
\label{fl1}}
\end{figure}

\begin{figure}[!ht]
\includegraphics[width=0.5\textwidth]{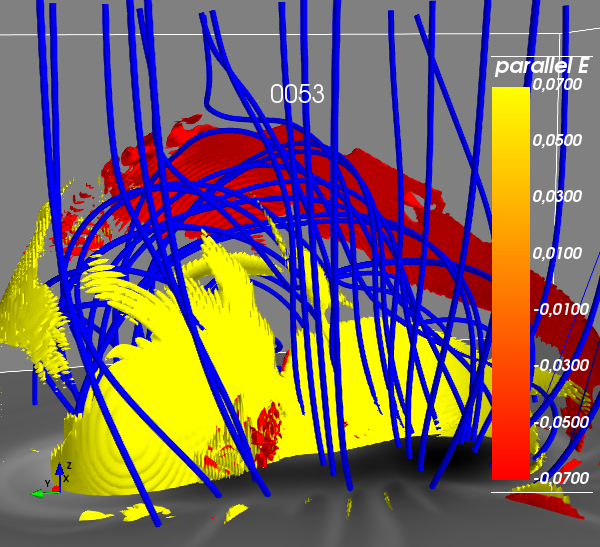}
\caption{The same as in figure 3, for the parallel electric field, and for snapshot 53. The eruption of the twisted magnetic field structure (flux rope) drives the emission of a blow-out jet.
\label{fl1_53}}
\end{figure}

With the focus of this study being on heating and acceleration in coronal 
active regions, we consider only the coronal part of the MHD 
simulation box. We consider two time-instances (termed snapshots in the following) out of the MHD simulation: (i) snapshot 30 (at $30\,$m from the simulation start), which shows a clearly  
shaped standard jet as a result of reconnection between the emerging and pre-existing magnetic fieldlines,
and (ii) snapshot 53 (at $53\,$m), at which a magnetic flux rope has been formed within the emergence region, and it erupts to drive the emission of an explosive ``blow out'' jet. It is to note that the time interval
between two MHD snapshots is $t_\mathrm{0}=85.7\ \mathrm{s}$, whereas the integration time of
the test-particles is of the order of $1\,$s, so there is no need
for e.g.\ interpolating in time direction between two subsequent 
MHD snapshots, we can consider one single snapshot and assume
that it does not evolve over the kinetic time-scale of interest. 
Fig.\ \ref{fl1} shows magnetic field-lines and iso-contours of the magnitude of the total electric field $|\mathbf{E}|$ (left panel, snapshot 30),
and iso-contours of the parallel electric 
field  $E_{||}$, including the sign (right panel, snapshot 30).
Both, the parallel and the total electric field 
are distributed along the main reconnection regions. These are the regions where current sheets are built up and the emerging field starts to reconnect with the pre-existing magnetic field. We find that the spatial distribution of the electric field reveals the formation of large scale structures (i.e.\ the current sheets) and their tendency to break up into fragments, in different degrees though,
with the parallel electric field clearly being more fragmented. Fig.\ \ref{fl1_53} shows the topology of the overall system at snapshot 53 of the MHD simulation. The magnetic field lines within the central emergence region are more twisted. This is due to the formation of a magnetic flux rope, owing to shearing and reconnection of field-lines above the polarity inversion line of the emerging bipolar region. As the flux rope rises, it starts to reconnect with the ambient field. Eventually, it erupts driving the onset of a helical blow-out jet. The spatial distribution of the electric field reveals two main regions where the electric fields are strong. One region is above the flux rope (red isosurface) and it shows the location of a large-scale current sheet at the interface between the erupting flux rope and the ambient coronal field. The second region is underneath the flux rope (yellow isosurface), and it 
marks
the location of the current sheet above the polarity inversion line where sheared field-lines reconnect to form the flux rope. 

\subsection{Statistics of the MHD electric field}

\begin{figure}[!ht]
\includegraphics[width=0.5\textwidth]{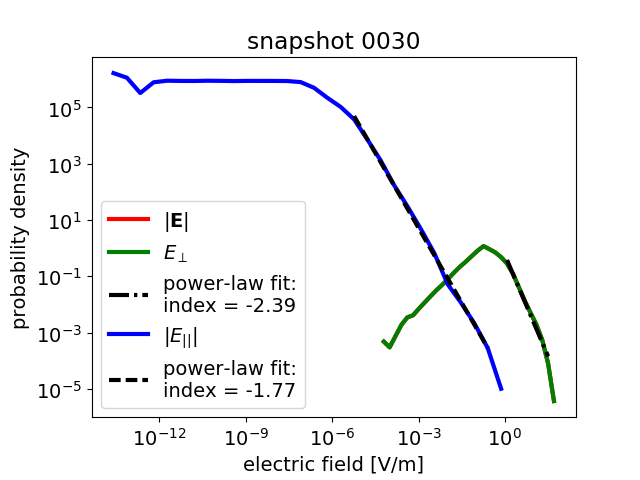}
\includegraphics[width=0.5\textwidth]{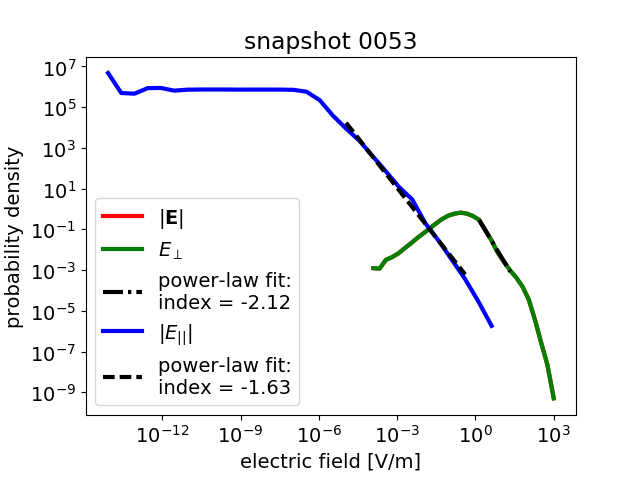}
\caption{
MHD simulations, snapshot 30 and 53, coronal part only: 
Distribution of 
the electric field from all coronal grid-points, 
for the magnitude of the total electric 
field, the perpendicular component (they practically coincide), and the parallel  component,
respectively.
The electric field is in units [V/m], and the mean Dreicer field
is $4.6\times 10^{-4}\,$V/m.
\label{Es}}
\end{figure}

Fig.\ \ref{Es} shows the histogram of the magnitude of the 
total electric field $|\mathbf{E}|$,
the parallel $|E_{||}|$ and the perpendicular $E_{\perp}$ components of the electric field,
determined from all coronal grid-points. They all show a power-law
tail with a roll-over at high values.
The power-law index of the fit is -1.8 for the parallel electric field,
and -2.4 for the total and perpendicular electric field in case of snapshot 30, and for snapshot 53 the values are similar, just that the total and perpendicular electric field attain larger values.
In any case, the parallel electric field
is 2 orders of magnitude smaller than the total electric field, which 
thus basically coincides with the perpendicular electric field.
Also, the parallel electric field shows a much more extended power-law tail
than the perpendicular and the total one. We thus conclude that power-law shaped distributions are inherent to the electric field and its components.
Similar results have also been found in MHD simulations of a decaying current sheet
	\citep[see Fig.\ 2 in ][]{Onofri06}
	and of a twisted coronal loop
	\citep[see Fig.\ 2b in ][]{Turkmani2006}.

The Dreicer field is given as 
\beq
E_{D}=  7\cdot 10^{-4} \frac{n}{10^9\, \mathrm{cm}^{-3}}
         \left(\frac{T}{10^7\, \mathrm{K}}\right)^{-1} 
         \frac{\ln \Lambda}{23.2} \, \frac{\mathrm{V}}{\mathrm{m}}
\eeq
\citep[e.g.][]{Holman1985},
so for typical coronal values used in the MHD simulations, $n=1.0\cdot10^{10}\,$cm$^{-3}$, $T=7.6\cdot 10^5\,$K, $\ln\Lambda=23.2$, we have the mean value
\beq
\langle E_{D}\rangle = 5.0\cdot10^{-4}\, \mathrm{V/m}
\eeq
with maximum value $2\times 10^{-2}\,$V/m and minimum value $2\times 10^{-6}\,$V/m.
It thus follows from Fig.\ \ref{Es} that the perpendicular
electric field is highly super-Dreicer almost everywhere, whereas
the parallel one attains highly super-Dreicer values only at a fraction
of the grid-points. The threshold $\pm 0.07$ chosen for the iso-contours
of the parallel electric field 
in Fig.\ \ref{fl1} (right panel) and Fig.\ \ref{fl1_53} corresponds to $140\langle E_{D}\rangle$,
implying that the parallel electric field is highly super-Dreicer 
in a wider region enclosing the main locations of reconnection and the outflow region.

\subsection{Statistics of the MHD energies}

\begin{figure}[!ht]
\includegraphics[width=0.5\textwidth]{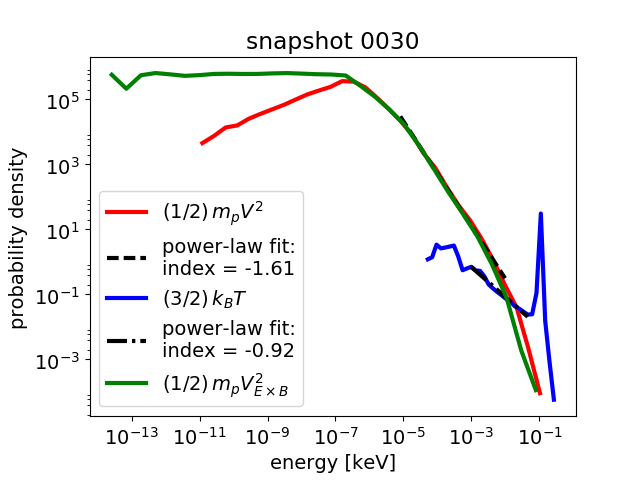}
\includegraphics[width=0.5\textwidth]{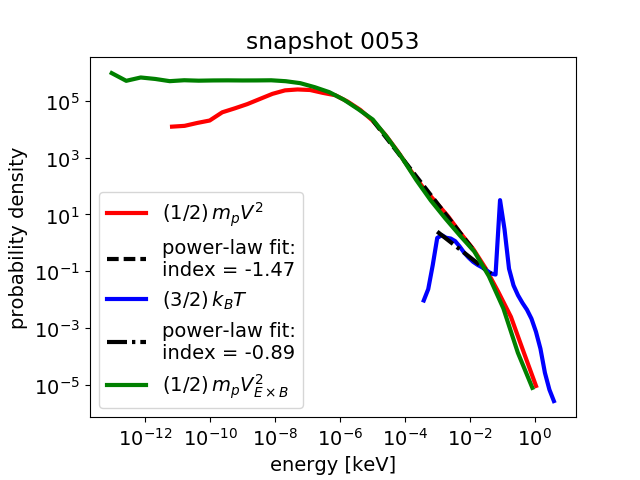}
\caption{
MHD simulations, snapshot 30 and 53, coronal part only: 
Distribution of the 
kinetic energy $(1/2) m_P V_{E \times B}^2$ of the E cross B drift velocity $V_{E \times B}$, of the kinetic energy $(1/2) m_P V^2$ of the MHD flow velocity $V$,
and of the thermal energy $(3/2) k_B T$, 
as determined from all the coronal grid-points.
\label{EcB}}
\end{figure}

Fig.\ \ref{EcB} shows the kinetic energy distribution of
the E cross B velocity, 
$\frac{1}{2} m_e (\mathbf{E}\times \mathbf{B}/B^2)^2$,
and the MHD flow velocity, $\frac{1}{2}m_e \mb{V}^2$, as calculated 
from all the coronal grid-points. 
For snapshot 30, the two distributions are very similar in shape --- except for the 
lowest energies ---, they show a power-law tail with index -1.61, and 
the highest energy reached is $0.1\,$keV. Also shown in the figure
is the MHD thermal energy distribution, $\frac{3}{2}k_B T$, which reaches a maximum value of $0.5\,$keV, has a power-law decay with index -0.92, and exhibits 
a clear peak near $0.1\,$keV. In case of snapshot 53, the situation is rather similar, just that roughly one order of magnitude larger energy values are reached.

\begin{figure}[!ht]
\includegraphics[width=0.5\textwidth]{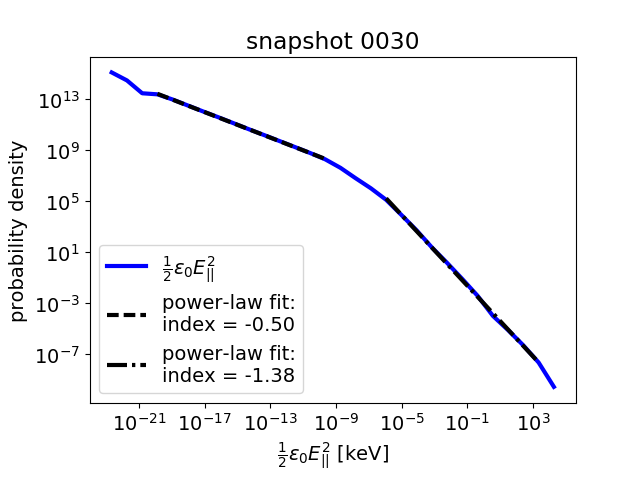}
\includegraphics[width=0.5\textwidth]{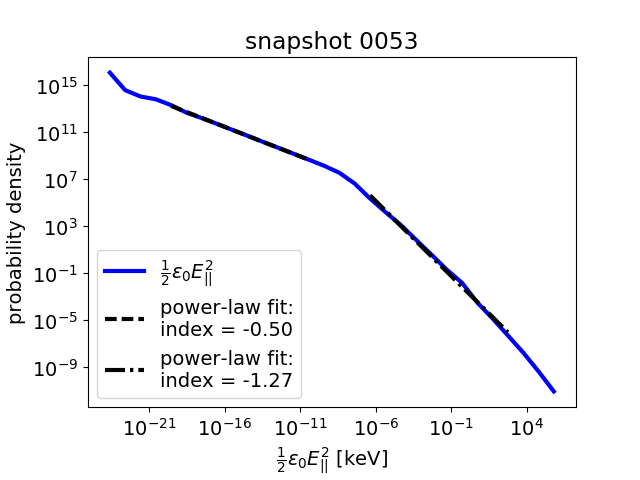}
\caption{
MHD simulations, snapshot 30 and 53, coronal part only: 
Distribution of the parallel electric field energy density 
$\frac{1}{2} \epsilon E_{||}^2$, 
as determined from all the coronal grid-points.
\label{eE}}
\end{figure}

Fig.\ \ref{eE} shows the distribution of the energy contained in the parallel electric
field,
\beq
W_{E_{||}} = \frac{1}{2} \epsilon E_{||}^2
\eeq
(with the permittivity $\epsilon \approx \epsilon_0$). For both snapshots considered, the distribution 
is of double power-law form, extending over many decades, with index 
-0.50 in the low energy part and with index -1.4 and -1.3, respectively,  in the high energy part, and where the maximum energy reached is about $10\,$MeV for snapshot 30,
and almost $100\,$MeV for snapshot 53.

It thus follows that all the MHD energies of interest exhibit power-law distributions, with relative low maximum value, of the order of the thermal energy (about $100\,$eV), with the exception of the energy contained in the parallel electric field that reaches the MeV regime. In the case of an emitted blow-out jet, we find that the MHD energies are generally one order of magnitude larger than at the standard jet.


\subsection{Spatial structure of the parallel electric field}

We now investigate the spatial structure of the parallel electric field, applying cluster analysis and calculating its fractal dimension.

\subsubsection{Cluster analysis}

We consider the magnitude of the parallel electric field $|E_\parallel|$
at all the coronal grid-points, and we a apply a threshold below
which $|E_\parallel|$ is set to zero. For the threshold we use the same 
value of $0.07$ as for the iso-contours of $E_\parallel$ in Fig.\ \ref{fl1}
and Fig.\ \ref{fl1_53}.

We define a cluster as a set of grid-points with (a) above-threshold value
of $|E_\parallel|$ at all the grid-points belonging to the cluster, and
(b) the cluster's grid-points are connected through their nearest 
neighbourhoods in 3D Cartesian coordinates. It follows that  
a cluster is surrounded by grid points with below-threshold $|E_\parallel|$. 
Defined in this sense, the set of all clusters is related to
the (above-threshold) iso-contours in Figs.\ \ref{fl1} and \ref{fl1_53}, just that 
the cluster analysis splits
the iso-contours into parts, the clusters, which are not connected 
through the nearest neighbourhoods of the grid-points. 
Each cluster is numbered uniquely,
and the grid-points belonging to it are marked correspondingly.

For snapshot 30, we find that there are 162 
clusters, and 
two of them are very dominant in spatial 
extent, one corresponding to the positive and one 
to the negative extended parallel electric field region 
in Fig.\ \ref{fl1}.

\begin{figure}[!ht]
\includegraphics[width=0.5\textwidth]{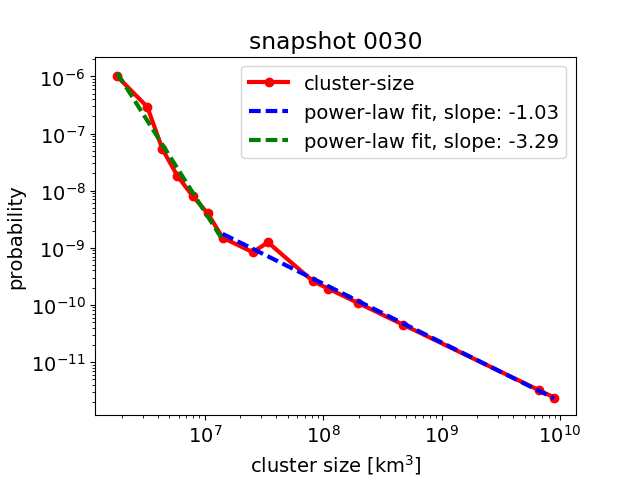}
\includegraphics[width=0.5\textwidth]{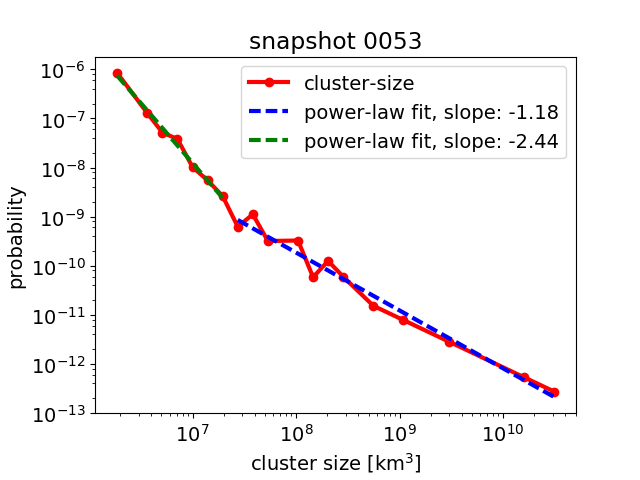}
\includegraphics[width=0.5\textwidth]{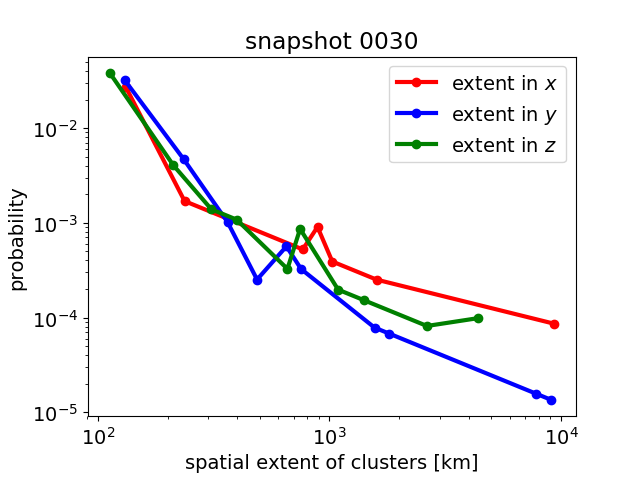}
\includegraphics[width=0.5\textwidth]{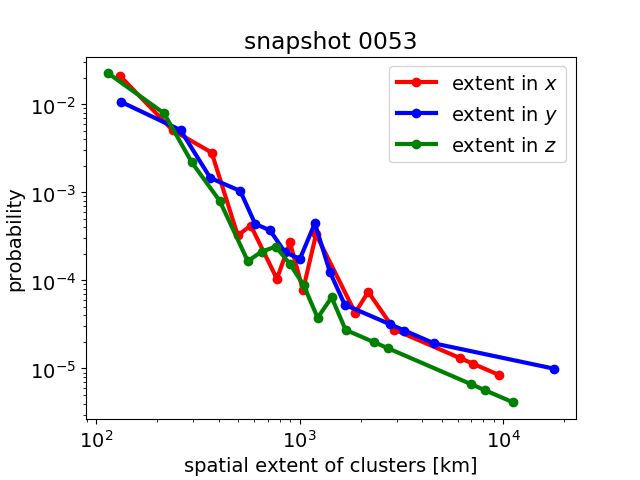}
\caption{
MHD simulations, snapshot 30 and 53, coronal part: 
Top: cluster-size distribution. Bottom: distributions of spatial
extent in the $x$-, $y$-, and $z$-direction.
\label{cl11}}
\end{figure}

For each cluster, we determine the cluster-size as the  
number of grid-points belonging to the cluster times 
the elementary grid-volume $\Delta x\,\Delta y,\Delta z$.
The distribution of cluster-sizes is shown in Fig.\
\ref{cl11}, there is a double power-law for both snapshots,
with a steep decay at small sizes with power-law index 3.3 and 2.4,
and a flatter power-law scaling at the larger sizes,
with index 1.0 and 1.2, for snapshot 30 and 53, respectively.
  
Fig.\ \ref{cl11} also shows the distribution of the linear 
cluster extents in the $x$, $y$, and $z$ direction. 
There again is a vague double power-law scaling, with the sizes being larger at the blow-out jet than at the standard jet.

\subsubsection{Fractal dimension}

\begin{figure}[!ht]
\includegraphics[width=0.5\textwidth]{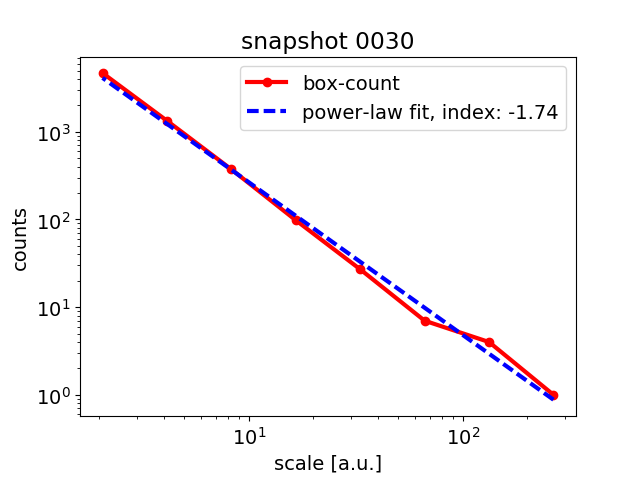}
\includegraphics[width=0.5\textwidth]{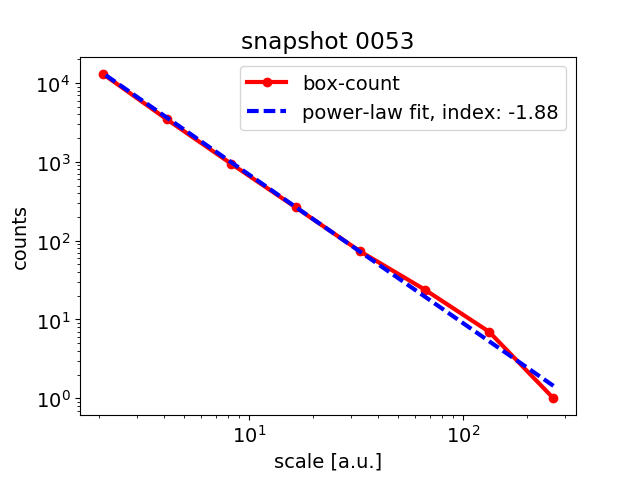}
\caption{Fractal dimension of the parallel electric field:
Scaling of the 3D box-counting algorithm, for snapshot 30 and 53.
\label{cl12}}
\end{figure}

To the same data as used in the cluster-analysis  
(the magnitude of the parallel electric field $|E_\parallel|$
at all the coronal grid-points, set to zero when 
below the threshold value of $0.07$), we apply a standard 
3D box-counting method in order to determine the fractal dimension $D_F$ of 
the region with above-threshold parallel electric field. Fig.\ \ref{cl12} shows
the scaling of the box-counts with the box-scale, there is 
a clear power-law scaling in the entire range, whose index, 
per definition of the box-counting method, equals the fractal dimension,
so we find $D_F = 1.7$ for snapshot 30 and $D_F = 1.9$ for snapshot 30. 

The regions of high parallel electric field can thus be interpreted 
as thinned out 2D-sheets, as it also corresponds to the visual impressions
that are given by 
Figs.\ \ref {fl1} and \ref {fl1_53}, and whereby the 'filling-factor' is higher at the blow-out jet compared to the time when the standard jet is emitted. 

After all, the spatial structure of the regions of strong parallel electric field can be characterized as fragmented and fractal, with the various size distributions exhibiting double power-law scalings.

\section{The evolution of the test-particles in the vicinity of the standard jet}

\begin{figure}[!ht]
\includegraphics[width=0.5\textwidth]{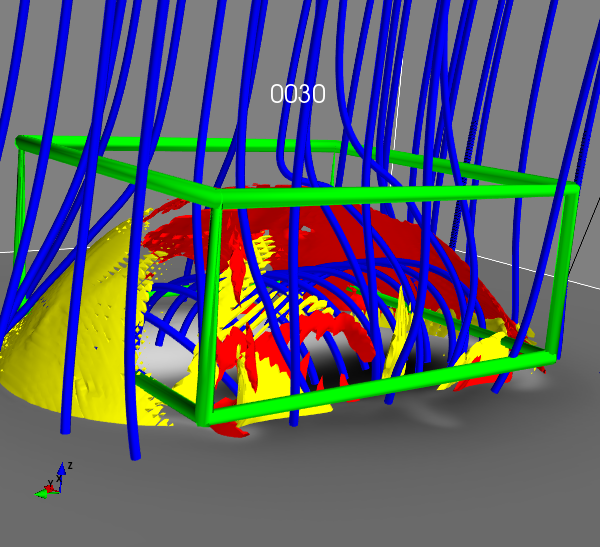}
\caption{
MHD simulations, snapshot 30, zoom into the coronal part: 
Magnetic field lines (blue), 
together with an iso-contour plot of the parallel electric 
field (red to yellow 3D-surfaces), 
for the 2 thresholds 
indicated by the color-bar in Fig.\ \ref{fl1}. At the bottom $x$-$y$-plane, 
the photo-spheric component $B_z$ is shown as a 2D filled contour plot. 
The 
region in which the spatial initial conditions are chosen is 
out-lined by a green cube. 
\label{fl31}}
\end{figure}

We first consider the energization of particles at the standard jet, snapshot 30.
The electric 
and magnetic fields from the MHD simulations are de-normalized to SI units and are not 
scaled further.
Electrons are considered as test-particles, and, 
if not mentioned otherwise, the standard integration time is $0.1\,$sec, 
and 100'000 particles are traced, in any case by using 
the relativistic guiding center approximation to the equations of motion,
see Sect.\ \ref{kinmod}.
The initial spatial position is uniform random in the region around 
the main reconnection region, as out-lined by the green cube
in Fig.\ \ref{fl31}, 
which basically contains the entire current sheet with all its fragments.
The initial velocity is random with Maxwellian 
distribution (i.e.\ Gaussian distribution of the velocity components), 
with temperature $\approx 9 \times 10^5\,$K. 
For each simulation, a set of 100 monitoring times has been predefined,
including the final time, at which the velocities and positions of the particles are 
monitored for the purpose of a statistical analysis to be done at equal
times for all the particles. Separate track is kept of the particles
that leave before the final time. 

\subsection{Acceleration}


\begin{figure}[!ht]
	\includegraphics[width=0.5\textwidth]{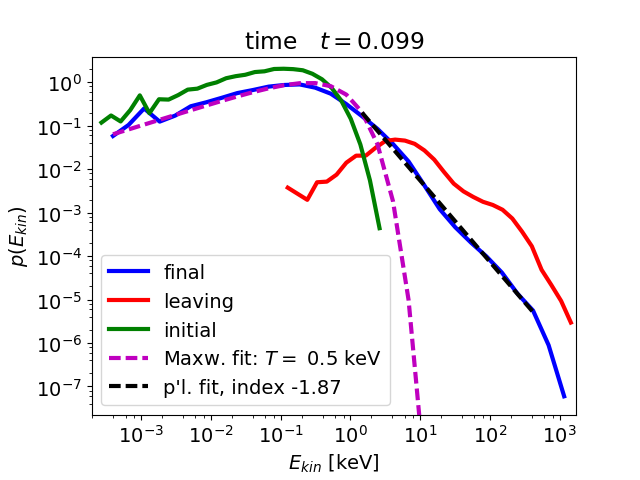}
	\includegraphics[width=0.5\textwidth]{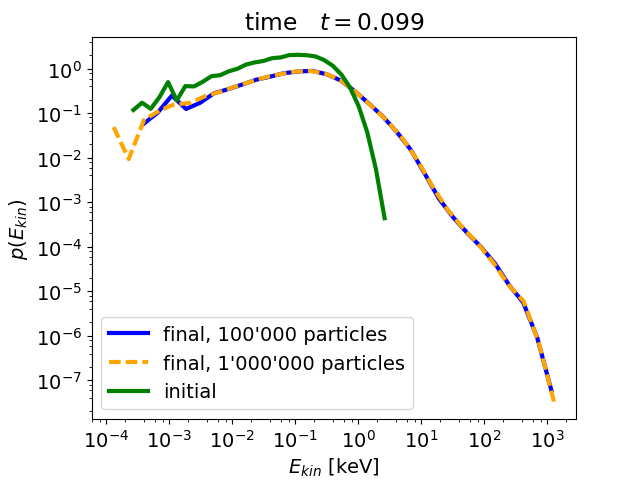}
	\includegraphics[width=0.5\textwidth]{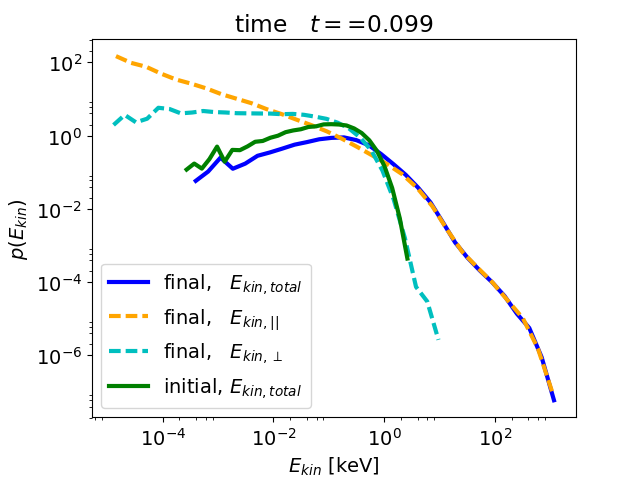}
	\includegraphics[width=0.5\linewidth]{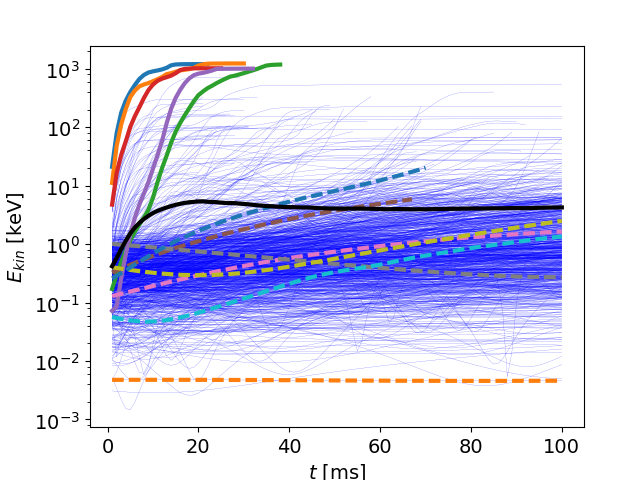}
	\caption{
		Snapshot 30: Kinetic energy distribution of electrons after $\approx 0.1\,$s
		(top left), 
		without 
		collisions, together with a fit at the low-energy, Maxwellian part and the 
		high energy, power-law part, the initial distribution, 
		and the 
		distribution of the leaving particles (for every particle at the time it 
		leaves).
		Top right: Comparison of the kinetic energy distributions from simulations
		with 100'000 and 1'000'000 test-particles, respectively.
		Bottom left: comparison of the total kinetic energy, the parallel,
		and the perpendicular kinetic energy at 0.1s.
        Bottom right: Kinetic energy of particles as a function of time (thin blue lines), with a few trajectories marked with bold lines (solid at high energies and dashed at low energies), together with the mean energy (solid black).
		\label{Ekin1}}
\end{figure}

Fig.\ \ref{Ekin1}, top left, shows the distribution of the kinetic energies
of the particles after $0.1\,$s, together with the initial distribution 
and the distribution of the leaving particles (as collected at the times
the individual particles leave). The final energy distribution is of 
Maxwellian shape at the low energies, and exhibits a slightly modulated
power-law tail.
The maximum energy reached is about $1\,$MeV, and
a power-law fit to the tail of the kinetic energy distribution 
yields an index of about -1.87. 

13\% of the 100'000 particle that are traced have left after $0.1\,$s,
and they
have energies in the same range than those
that stay inside, with a modulated power-law tail
that is 
steeper though, with index -2.98 at the highest energies (the fit is not shown).

In fig.\ \ref{Ekin1}, bottom left, we separately show the final
total, parallel, and perpendicular kinetic energy at $0.1\,$s. Obviously,
the power-law tail in the total kinetic energy is inherited from 
the parallel kinetic energy, there is essentially no energization 
in the perpendicular direction, as expected from Fig.\ \ref{EcB},
with the energy in the E cross B velocity having a maximum value 
of only $0.1\,$keV.
Thus, an important conclusion is that acceleration is 
acting 
exclusively in the parallel direction. 

In order to check how reliable the statistical sample of 100'000 particles
is, we performed a simulation with 10 times more particles (1'000'000),
and from Fig.\ \ref{Ekin1}, top right, 
it can be seen that the kinetic energy 
distribution does not change when using a substantially higher number 
of particles, in particular also the maximum energy reached
remains unchanged. We thus will use throughout 100'000 particles
in the simulations presented.

Fig.\ \ref{Ekin1}, bottom right, shows the kinetic energy of the particles as a function of time, with a few low energy and a few high energy particles marked, together with the mean energy. The high energy particles have the tendency to reach their final energy in a fast and single step, with a time-scale of the order of some tens of ms. The low energy particles evolve on a much slower time-scale, of the order of seconds, during which the energy gradually increases; in particular, they show rather a slow drift motion than a classical random walk.


\subsection{Heating}

\begin{figure}[!ht]
\includegraphics[width=0.5\textwidth]{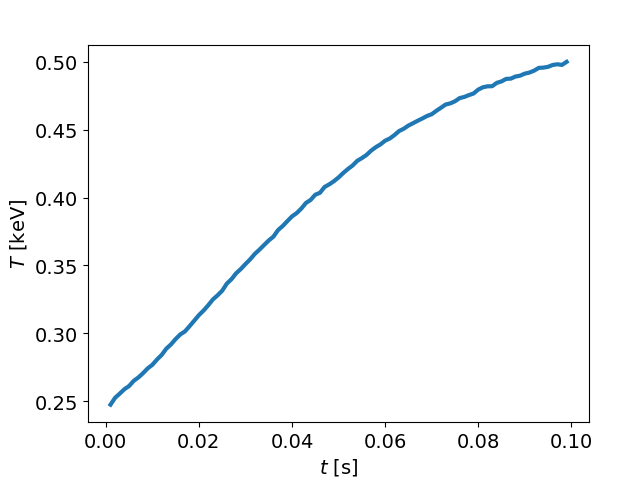}
\includegraphics[width=0.5\textwidth]{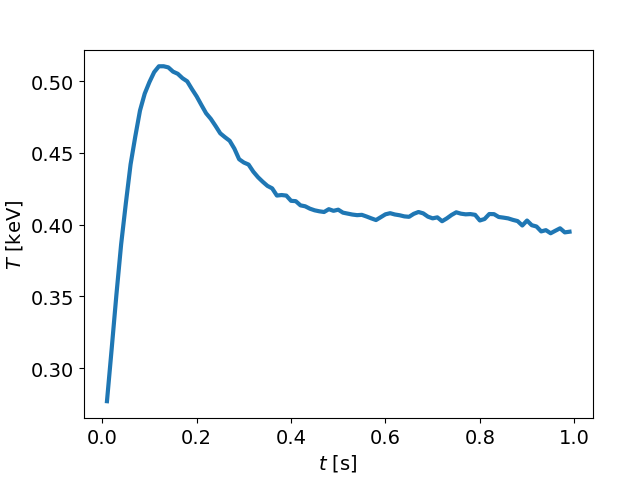}
\caption{
Snapshot 30: Temperature as a function of time, as determined through 
the fit of a Maxwellian in the low energy range of the kinetic 
energy distribution.
\label{Tt}}
\end{figure}

For the particles that stay inside, 
the Maxwellian shape of the energy distribution is well 
preserved at low energies, 
and there is heating from the initial $0.24\,$keV to $0.50\,$keV after 
$0.1\,$s, 
as the Maxwellian fit in Fig.\ \ref{Ekin1}, top left, reveals. 
Fig.\ \ref{Tt} shows the temperature as a function of time,
as estimated by Maxwellian fits to the kinetic energy distribution in the low energy part. 
The temperature increases linearly with time until $0.05\,$s and then 
starts to turn over, reaches a peak value of $0.50\,$keV at $0.1\,$s, and 
finally the heating process saturates for times larger than  
roughly $0.5\,$s with a temperature of $0.40\,$keV. The decrease from 
$0.5$ to $0.4\,$keV may be attributed to the loss of a fraction of the higher energy bulk particles.

The energy distribution of the leaving particles 
shows a functional 
form at low energies (between $0.1$ and $10\,$keV) that is reminiscent of a Maxwellian, and 
a respective fit reveals a temperature of about $13.3\,$keV
(see Fig.\ \ref{Ekin1}, top left, the fit itself is not shown). Although the statistics is not 
very good, we can interpret these particles as belonging to a super-hot
population. It is to note though that the energies are monitored 
at different times for each particle, so the distribution is asynchronous.

\subsection{Longer and shorter times}

\begin{figure}[!ht]
	\includegraphics[width=0.5\textwidth]{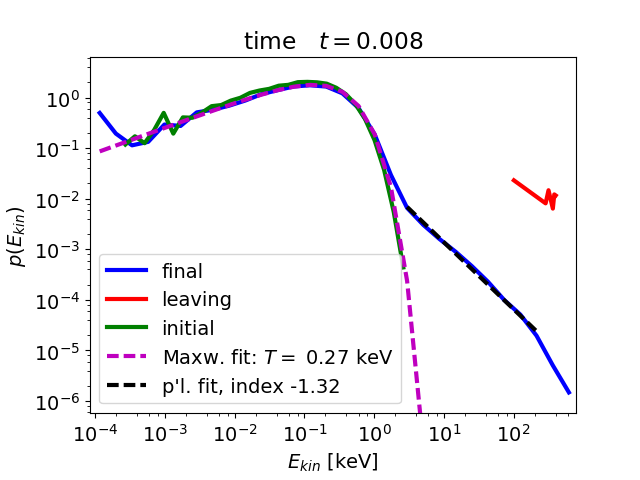}
	\includegraphics[width=0.5\textwidth]{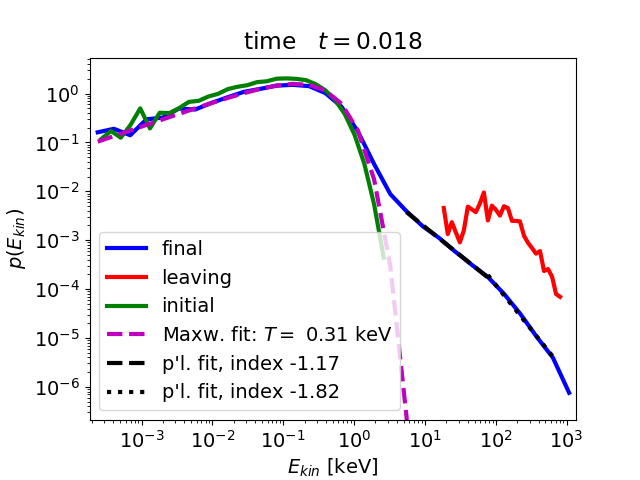}
	\includegraphics[width=0.5\textwidth]{Ekin_0_1.png}
	\includegraphics[width=0.5\textwidth]{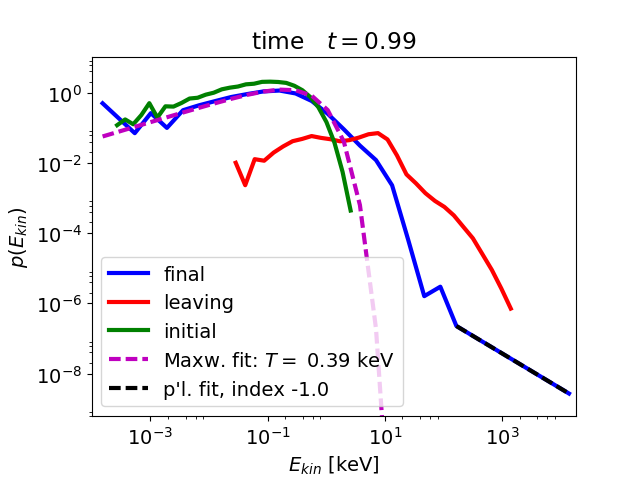}
	\caption{
		Snapshot 30: Kinetic energy distribution of electrons after $\approx 0.01\,$s
		(top left), $\approx 0.02\,$s (top right), $\approx 0.1\,$s (bottom left) and $\approx 1.0\,$s (bottom right), without 
		collisions, together with a fit at the low-energy, Maxwellian part and the 
		high energy, power-law part, the initial distribution, 
		and the 
		distribution of the leaving particles (for every particle at the time it 
		leaves).
		\label{Ekin1_times}}
\end{figure}

Considering times longer than $0.1\,$s, we find that
at a final time of $1.0\,$s, 57\% of the particles have left the 
system.
The kinetic energy distribution of the particles is shown
in Fig.\ \ref{Ekin1_times}, bottom right.
The tail of the distribution for the particles that remain inside has now a clear power-law part only at the 
highest energies, with index $-1.0$, much smaller than the one at $0.1\,$s.
The intermediate to high energy part does though not show a clear scaling anymore,
the statistics has become poor  due 
to the large number of particles that has left. It is to note though that
the highest energy reached
of about $20\,$MeV is much larger  
in comparison to the time $0.1\,$s. At the low energies, 
the particles are heated to a temperature of $0.40\,$keV, 
which is below the peak temperature reached at $0.1\,$s.

The leaving particles have a modulated power-law tail, with index $-2.73$
at the highest energies, which is close to the index seen at $0.1\,$s, and the low energy part is now 
closer to a Maxwellian shape, with a temperature of $7.5\,$keV, about one 
half of the temperature at $0.1\,$s (both fits are not shown in the figure).

Considering times shorter than $0.1\,s$, we find that at 
$t\approx 0.01\,$s a power-law tail has already been formed with index $-1.32 $ (Fig.\ \ref{Ekin1_times} top left), and at $t\approx 0.02\,$s a double power-law appears with index $-1.82$ in the high energy part (Fig.\ \ref{Ekin1_times} top right). The power-law tail thus gets steeper in the initial phase, and then flattens for larger times.

\subsection{The effect of  collisions}

\begin{figure}[!ht]
\includegraphics[width=0.5\textwidth]{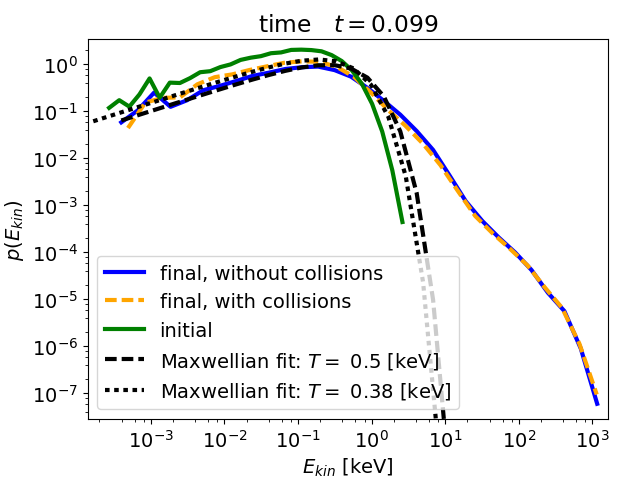}
\caption{
Snapshot 30: Kinetic energy distribution of electrons after 0.1 sec, without 
collisions (blue) and with collisions (orange), together with 
the initial distribution (green).
\label{Ekin_coll_rel}}
\end{figure}

We consider collisions with background electrons of the same temperature
as the initial temperature of the test-particles (see Sect.\ \ref{kinmod}). As Fig.\ 
\ref{Ekin_coll_rel} shows, collisions play a
role at low energies only, as expected, 
they reduce the efficiency of the heating process, cooling down the 
electrons towards their initial 
temperature, the temperature reached at $0.1\,$s with collisions 
is $0.38\,$keV, 
compared to $0.50\,$keV in the case without collisions. The 
cooling down of the test-particles corresponds of course to 
a heating of the background population, which is not taken 
into account in our modelling approach.

\subsection{Particle orbits}

\begin{figure}[!ht]
\includegraphics[width=0.5\textwidth]{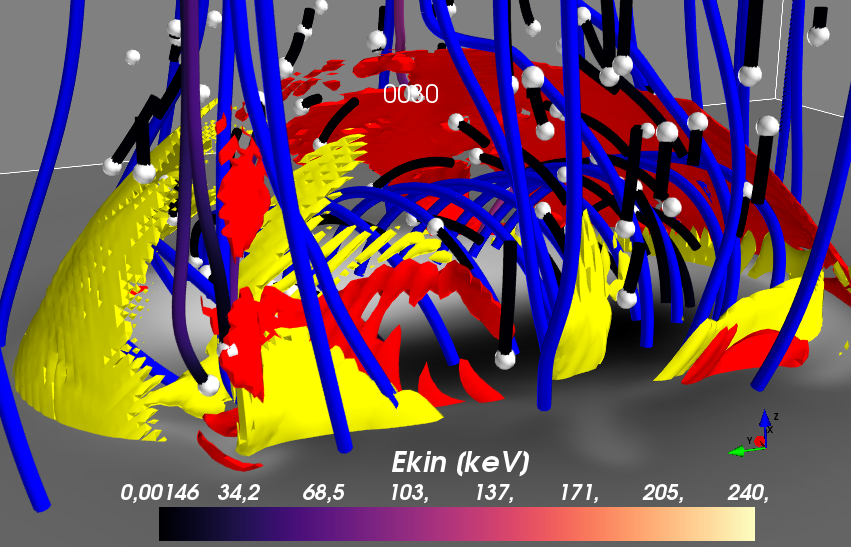}
\includegraphics[width=0.5\textwidth]{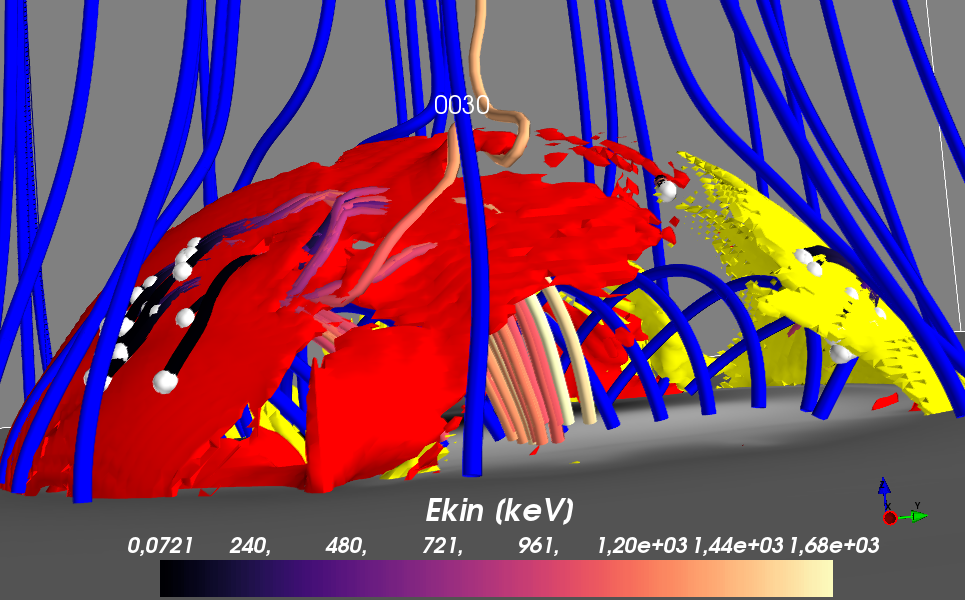}
\caption{
MHD simulations, snapshot 30, zoom into the coronal part: 
Magnetic field lines (blue), 
together with an iso-contour plot of the parallel electric 
field (red to yellow 3D-surfaces), for the 2 thresholds as in Fig.\ \ref{fl1}. 
A sample of 40 particle orbits is shown, as tubes, 
with the color indicating the value of $E_{kin}$ in keV according to the 
color-bar,
and with the initial conditions marked with white spheres. 
In the left panel, the 40 orbits are randomly chosen, and in the right panel 
the orbits of the
40 most energetic particles are shown.
\label{fl3}}
\end{figure}

Fig.\ \ref{fl3}, left, shows particle orbits for 40 randomly chosen 
particles. The particle energy is indicated by the color of the orbits.
Being randomly chosen, the particles mainly belong to the population
that is heated or moderately accelerated, and there is no pattern 
discernible concerning any preferences.

Fig.\ \ref{fl3}, right, shows the orbits of the 40 most 
energetic particles, and now there is clearly visible a preference
in the initial conditions, there are two regions from where the energetic particles
originate, one close to the region of strong positive and one close to the
region of strong negative parallel electric field. Also, 
most particles move some distance along the region of high electric 
field and 
pass through it at some point, whereby their energy increases
strongly, and after which 
they 
leave the simulation cube, mostly, but not all, through the bottom plane, towards
the photosphere. The energetic particles thus undergo just one acceleration event, 
in accordance with the picture given in Fig.\ \ref{Ekin1}, bottom right,
and hence the acceleration process is of a single and not a multiple nature.

\section{Particle dynamics during the blow-out jet emission}

\begin{figure}[!ht]	
\includegraphics[width=0.5\textwidth]{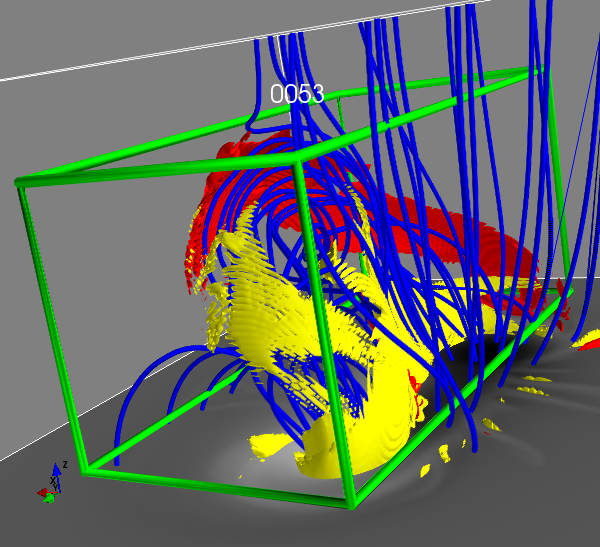}
\caption{
MHD simulations, snapshot 53, zoom into the coronal part: 
Magnetic field lines (blue), 
together with an iso-contour plot of the parallel electric 
field (red to yellow 3D-surfaces), for the 2 thresholds as in Fig.\ \ref{fl1_53}. 
The green cube outlines the region from which the initial conditions are 
chosen.
\label{fl53}}
\end{figure}

\begin{figure}[!ht]
\includegraphics[width=0.5\textwidth]{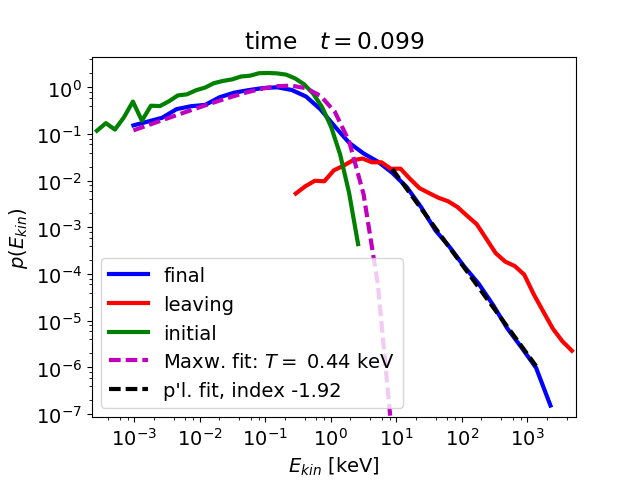}
\includegraphics[width=0.5\textwidth]{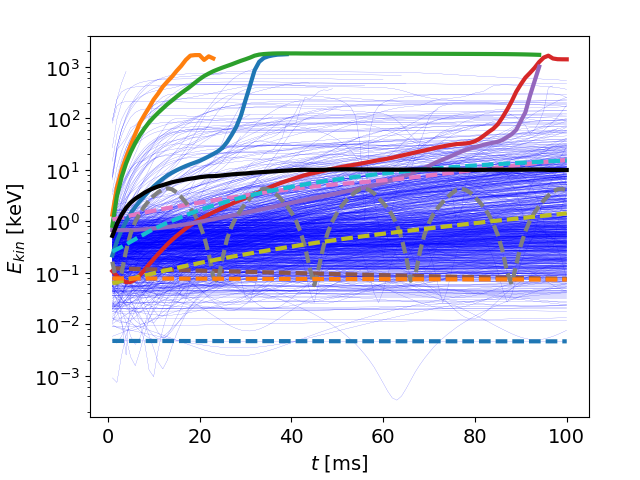}
\caption{
Snapshot 53: Left: Kinetic energy distribution of electrons after 0.1 sec,
without collisions, 
together with a fit at the low-energy, Maxwellian part and the 
high energy, power-law part, the initial distribution, 
and the 
distribution of the leaving particles (for every particle at the time it 
leaves). --- Right: Kinetic energy as a function of time for the test-particles (thin blue lines), with a few high energy (solid) and low energy (dashed) particles marked with different colors, together with the evolution of the mean value (solid black).  
\label{Ekin53}}
\end{figure}

We now turn to snapshot 53, at which a well developed blow-out jet has been formed.
Fig.\ \ref{fl53} shows the magnetic configuration, the structure
of the parallel electric field, and the region from which the initial
conditions are chosen. Fig.\ \ref{Ekin53} shows the kinetic 
energy distribution at $t\approx 0.1\,$s, there is heating to a temperature of $0.44\,$keV at the low energies,
and there is also acceleration, with the power-law tail having index -1.92. Both the temperature and the power-law index are close to the values found in case of the standard jet, just the highest
energy reached ($2\,$GeV) is now twice as large.

The leaving particles show a clear power-law tail with index -1.8, and they are heated to a temperature of $17.3\,$keV. The temperature thus is close to the one reached at the standard jet, the power-law tail though is much flatter now.

Fig.\ \ref{Ekin53}, right, shows the kinetic energy of the particles as a function of time. Similar to the case of the standard jet, the low energy particles drift slowly upwards in energy, while the high energy particles undergo basically one acceleration event (one-step or single acceleration).

The above results indicate that the particle energetics on the kinetic level 
are very similar in the standard and the blowout jet cases.  

\section{Nature of transport in energy space}

We now turn to the question about the nature of transport in energy space, in view of the results reported in the previous sections that are based on the test-particle approach.

\subsection{The classical Fokker-Planck equation approach}

A first and classical candidate for a statistical transport model is the Fokker-Planck equation, which in energy space writes as
\beq
\frac{\p f}{\p t}= \frac{\p}{\p E}\left[ \frac{\p (D f)}{\p E}  - F f\right]
-\frac{f}{\tau_{esc}}
\eeq
with $f$ the probability density function of the kinetic energy,
$D$ the diffusion and $F$ the convection coefficient, and $\tau_{esc}$
the escape time.
In this approach, the basic step is the determination of the two transport coefficients, $D$ and $F$, which we derive here from the test-particle simulation data.

In the following, we denote with $E_{kin,j}(t_{k})$ the kinetic energy of the particle with index $j$ at the predefined, equally spaced, monitoring times $t_k$ ($k=1,...,100$, $t_{100}$ is the final time).
In order to determine the energy- and time-dependent transport 
coefficients from the test-particle simulation data, we follow 
\cite{Ragwitz2001}, whose definition is based on the time-dependent energy differences
\beq
\epsilon_j(t_k) := E_{kin,j}(t_{k+h})-E_{kin,j}(t_{k})
\label{eq:epsilon}
\eeq
with $h$ the lag index, and usually we set $h=1$.

An estimate of the energy-dependence of the 
transport coefficients, for a given 
time $t_k$, is made by first prescribing bins 
along the $E_{kin}$-axis, with mid-points $E_i$ ($i=1,...,n$), and then considering 
$E_{kin,j}(t_{k+h})-E_{kin,j}(t_k)$
a function of $E_i$ if 
$E_{kin,j}(t_k)$ lies in the bin $i$.
The functional form of the transport coefficients, defined as
\beq
D\left(t_k,E_{i}\right) 
= \frac{1}{2(t_{k+h}-t_{k})}\left\langle 
\left(E_{kin,j}(t_{k+h})-E_{kin,j}(t_{k})\right)^2  
\right\rangle_j
\left(t_k,E_i\right)
\eeq
and
\beq
F\left(t_k,E_{i}\right) 
= \frac{1}{(t_{k+h} - t_{k})}
\left\langle 
E_{kin,j}(t_{k+h}) - E_{kin,j}(t_{k})
\right\rangle_j
\left(t_k,E_i\right)
\eeq
(see \cite{Ragwitz2001})
can then be determined by applying binned statistics, i.e.\ by calculating  the 
mean values
for each energy-bin separately at a given time instance $t_k$.

\begin{figure}[!ht]
\includegraphics[width=0.5\textwidth]{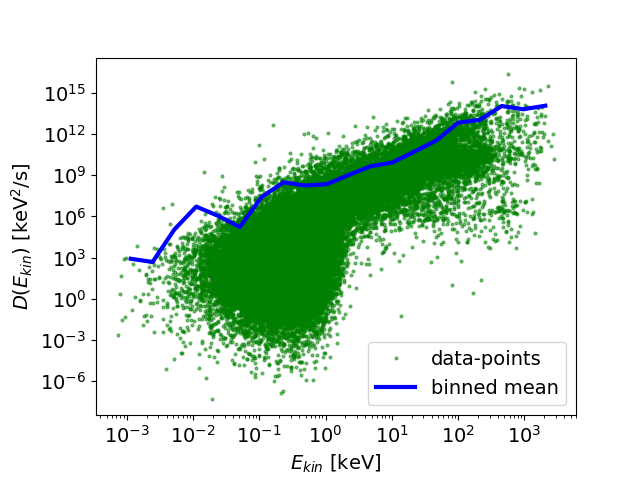}
\includegraphics[width=0.5\textwidth]{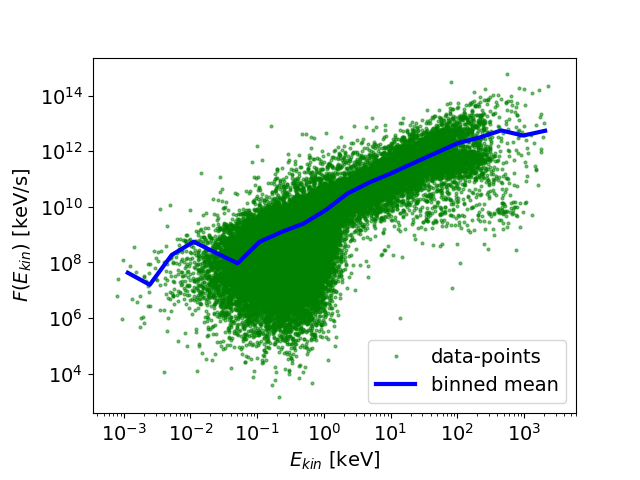}
\caption{
Snapshot 53: Diffusion coefficient $D$ (left),
and convective coefficient $F$ (right), 
as a function of the kinetic energy, at time $t \approx 0.1\,$s,
together with the data-points on which the binned statistics is based.
\label{DDD}}
\end{figure}

Fig.\ \ref{DDD} shows the diffusion
and drift coefficient for snapshot 53 at $t=t_{100}\approx 0.1\,$s, 
together with the data-points that form the sample of the binned statistics
($\epsilon_j(t_{100})^2/2(t_{k+h} - t_{k})$ and $\epsilon_j(t_{100})/(t_{k+h} - t_{k})$, respectively, as a function of $E_{kin,j}(t_{100})$). The binned mean for both coefficients exhibits quite a clear power-law functional form, yet the spread of the data points around the mean values is large, namely several orders of magnitude, which is a first hint that the estimate of the transport coefficients is problematic. 

To clarify the situation further, we show in Fig.\ \ref{pwcharfun} the histogram $p(\epsilon)$ of the energy increments $\epsilon_j$, which follows a double power-law distribution, with index $-1.97$ at the highest energies. It then follows that the drift coefficient $F$, as a mean values of the increments, is not representative of the scale-free data, and the diffusion coefficient $D$, as a variance of the increments, is ill-defined. This result, namely that the Fokker-Planck formalism breaks down, has been found also for the cases of strong turbulence
\citep{Isliker2017a} and of turbulent reconnection \citep{Isliker2017}.

\subsection{Fractional transport equation}

The power-law tail of the distribution of energy increments implies that the particle dynamics is anomalous, with occasionally large energy steps being  made, the particles perform Levy-flights in energy space when their dynamic is interpreted as a random walk.
In \cite{Isliker2017a}, we have introduced a formalism for a fractional transport equation (FTE) that is able to cope with this kind of non-classical dynamics. In this approach the distribution of energy increments is interpreted as a symmetric stable Levy distribution $p_\epsilon(\epsilon;\alpha,a)$, which are defined in Fourier space (characteristic function, $\epsilon\to k$) as $\hat p_\epsilon(k;\alpha,a)=\exp(-a |k|^\alpha)$, and which have a power-law tail in energy-space, 
$p_\epsilon(\epsilon;\alpha,a) \sim \epsilon^{-(1+\alpha)}$
\citep[see e.g.][]{Hughes1995}. The energy increments are sampled over constant time-intervals $\Delta:=t_{k+h} - t_k$, which has as a consequence that the temporal part of the FTE is non-fractional.

The FTE has the form \citep[for details and its derivation see][]{Isliker2017a}
\beq
\frac{\p f}{\p t} = \frac{a}{\Delta} D^\alpha_{|E|} f
-\frac{f}{\tau_{esc}}
\eeq
with $ D^\alpha_{|E|} $ the symmetric Riesz fractional derivative of order $\alpha$ (defined in Fourier space as $\mathcal{F}(D^\alpha_{|E|}f)=-|k|^\alpha \hat f$), $a$ the constant of the Levy stable distribution that is related to the width of the distribution of increments, and $\Delta$ the applied time-step in monitoring the particles' energy increments, defined just above.  

\begin{figure}
	\centering
	\includegraphics[width=0.4\textwidth]{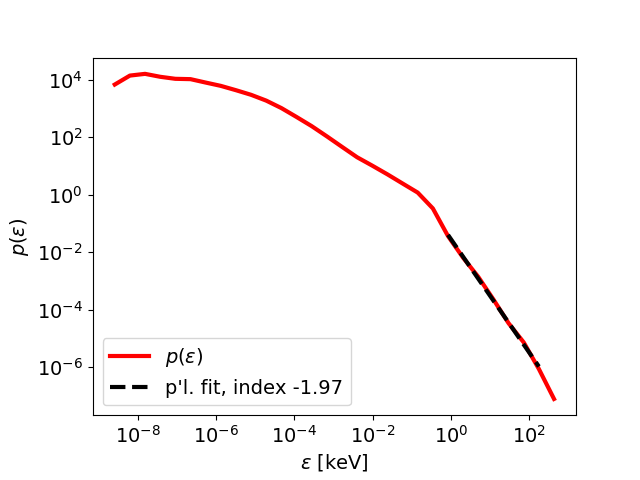}
	\includegraphics[width=0.4\textwidth]{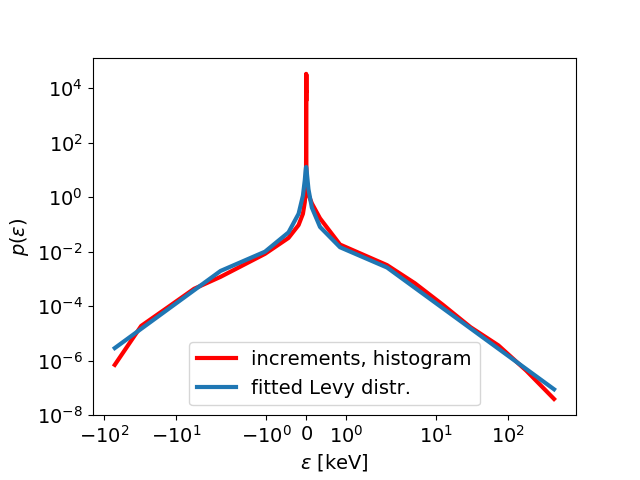}
	\includegraphics[width=0.4\textwidth]{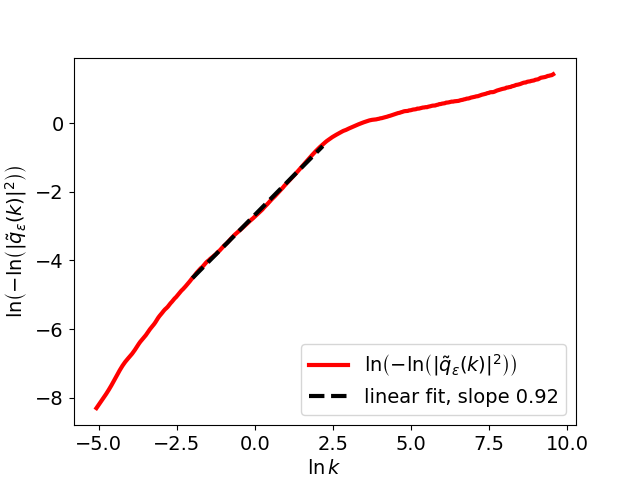}
	\caption{
		Snapshot 53, $t\approx 0.1\,$s: (a) Distribution of energy increments. (b) Two-sided distribution of energy increments, together with the fitted stable Levy distribution. 
		(c) Characteristic function estimate.}
	\label{pwcharfun}
\end{figure}

With $\Delta$ already  given, we still need to determine two parameters, $\alpha$ and $a$. A first way to infer $\alpha$ is through the index of the power-law in the tail of the distribution of energy increments, which yields $\alpha=0.97$, see Fig.\ \ref{pwcharfun}. A second way to determine $\alpha$ and also $a$ is through the characteristic function method \citep[e.g.][]{Borak2005,Koutrouvelis1980}, as described in \cite{Isliker2017a}, which gives $\alpha = 0.92$ and $a=0.03$, see Fig.\ \ref{pwcharfun}. 
A third way to infer the parameters is by applying a maximum likelihood estimate \citep[e.g.][]{Borak2005}, which for financial data is known to be the most precise method, in our application though we find large deviations when comparing to the index of the power-law tail in the increments, since the method yields $\alpha=1.24$, $a=3.2$.    
The fit of the Levy stable distribution is shown in Fig.\ \ref{pwcharfun},
with the parameters estimated through the characteristic function method, 
and it indeed
is in good agreement with the distribution in the power-law tail.

\begin{figure}[h]
	\centering
	\includegraphics[width=0.4\linewidth]{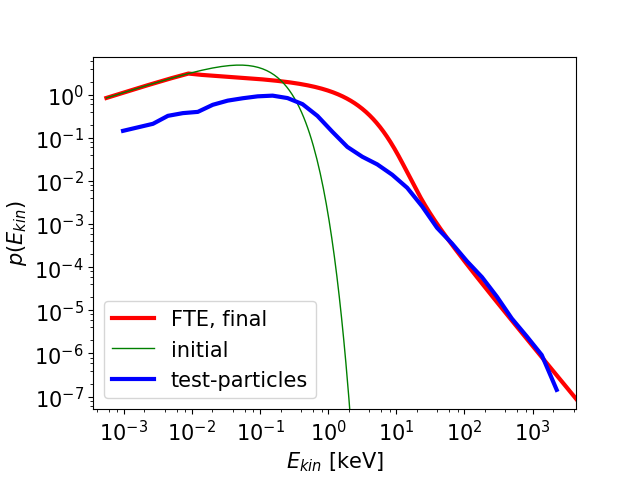}
	\caption{Snapshot 53: Kinetic energy distribution at initial and final time ($t\approx 0.1\,$s) from the test-particle simulations, together with the solution of the FTE at final time.}
	\label{fig:ekinfract}
\end{figure}

For the numerical solution of the FTE, we use the Gr\"unwald Letnikov definition of fractional derivatives \citep[see e.g.][]{Kilbas2006}, in the matrix formulation of \cite{Podlubny2009}, and in order to allow for a logarithmically equi-spaced grid on the energy axis, we make use of the formulation for non-equi-distant grid-points in \cite{Podlubny2013}. Time-stepping is done by the backward Euler method. The solution of the FTE at the same final time as for the test-particles is shown in Fig.\ \ref*{fig:ekinfract}, there is very good coincidence with the distribution function from the test-particle simulation 
for the entire power-law tail.
The FTE is thus an adequate transport model for the acceleration process of the high energy, power-law distributed particles.   
We note that modeling transport in energy space isolated from the simultaneous spatial transport, despite being of valuable interest, is a simplification, for a full understanding of the dynamics a model for the combined transport in energy and position space  is needed, which though first needs to be developed and which seems a nontrivial problem for our case here of doubly anomalous transport (see also the next section).

\section{Spatial diffusion}

We define the spatial mean square displacement (MSD) as 
\beq
\left\langle \rho_j^2 \right\rangle := \frac{1}{2(t_{k+h}-t_{k})}\left\langle \left(\mb{x}_{j}(t_{k+h})-\mb{x}_{j}(t_{k})\right)^2  \right\rangle_j
\eeq
i.e.\ we formally follow the definition of the displacements in energy, Eq.\ (\ref{eq:epsilon}), which explicitly allows for a time-dependent MSD. 

\begin{figure}
	\centering
	\includegraphics[width=0.45\linewidth]{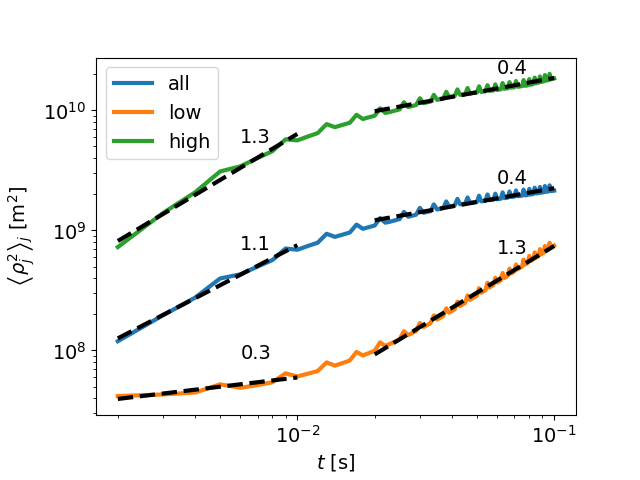}
	\includegraphics[width=0.45\linewidth]{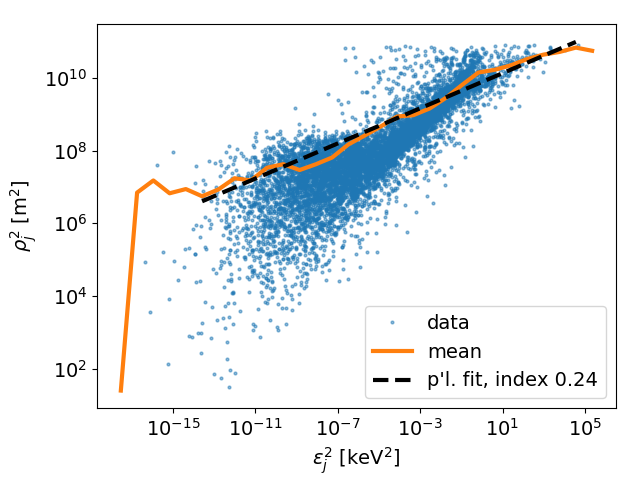}
	\caption{(a) Instantaneous mean square displacement in space as a function of time for all the particles ('all'), the high ('high') and low ('low') energy particles, together with power law fits (dashed lines), with the numbers indicating the power-law indexes of the fits. (b) Scatter plot of the squared spatial increments $\rho_j^2$ as a function of the squared energy increments $\epsilon_j^2$, together with the binned mean and a power-law fit to the latter.}
	\label{fig:rho2eps2}
\end{figure}

The MSD is shown in Fig.\ (\ref{fig:rho2eps2}). The particles are moderately super-diffusive in the initial phase up to $0.01\,$s, and then they turn over and become clearly sub-diffusive. Also, we have separated the particles into two populations, one that reaches high energies and forms the tail of the distribution, $E_{kin}>20\,$keV, and one that corresponds to the low energy, bulk population, $E_{kin}<20\,$keV. The high energy population shows a very similar behavior as the entire sample of particles, it just is more clearly super-diffusive at small times. They low energy particles, on the other hand, are sub-diffusive at small times, and then turn to super-diffusive at large times, they thus show the inverse behavior of the high energy particles. 

Fig.\ (\ref{fig:rho2eps2}) also shows the increments $\rho_j^2$ as a function of the corresponding energy increments $\epsilon_j^2$. There is a correlation between the two kind of increments, with a power-law shaped functional dependence, the spread though of the data points around this power-law is rather large. 
There thus is a trend for large energy increments to be associated with large spatial increments.

%

\section{Discussion and conclusions} \label{sumdisc}

Using a 3D resistive MHD code, \cite{Archontis13} simulated the emergence and eruption of solar magnetic fields interacting with a unipolar magnetic field. Their analysis was concentrated on the macroscopic parameters during the formation and emission of a standard and a more explosive blow-out jet. A key element in their analysis was the formation and the evolution of the reconnecting current sheet at the base of the jet, where the emerging magnetic flux meets the ambient magnetic field. We use their numerical results for the electromagnetic fields to analyze the heating and acceleration of particles in the vicinity of the reconnecting current sheets and the characteristics of the particles escaping from the energy release volume. The  interaction of the emerging flux with the ambient magnetic field is evolving on scales of tenths of minutes and ignites the standard jet 30 minutes after the initiation of the numerical experiment. The blow out  jet reaches its peak of activity about 20 minutes later. The acceleration time   of the electrons, in the presence of the electromagnetic fields of the MHD simulations, is much faster, their energy distribution reaches saturation in less than one second.  In the analysis presented here, we study the statistical properties of the electric field at two particular times, $t=30\,$m and $t=53\,$m, i.e.\ during the formation of the standard and the blow out jet, and we investigate the kinetic evolution of electrons at these two times.

The electric fields spontaneously develop a fragmented and fractal structure 
in the spatial vicinity of the reconnecting current sheet, 
and the clusters constituting the fractal obey 
power law size distributions. The probability distribution of the electric field and its parallel component, the parallel electric field's energy density, and the MHD energies all exhibit power law tails.

The electrons interacting with the fragmented electric fields  are heated  and accelerated near the base of the jet at the times where the standard and blow out  jets are formed. 
The highest energies reached are larger by roughly a factor of 2 at the blow out jet than at the standard jet, and in both cases the power law index of the tail of the energy distribution is evolving in time. The heating of the particles at low energies is rather gradual, reaching saturation at a temperature of $4.5\,$MK (see the observations reported by \cite{Bain09}).
The escaping high energy particles form a super-hot population with temperature $150\,$MK, and with a power law tail \citep{Glesener12, Glesener2018, Chen13}. 

The acceleration is a fast one step process, lasting typically about $10\,$ms, there is no random walk like behavior, so there is no Fermi process operating, contrary to the widely assumed scenarios reported in the current literature \citep{Drake06, Kowal11, Drake13, Lazarian15, Guo15} for the acceleration of particles in fragmented current sheets. 
The heating mechanism also does not follow a random walk process but is rather of a systematic drift nature.

The acceleration of the electrons is exclusively in the parallel direction and is clearly and solely caused by the parallel electric field, 
whose statistical properties in the vicinity of the unstable
current sheet
thus play the key role in the characteristics of the heated and accelerated particles.
The power law distributions of both, the strengths of the parallel electric field and the sizes of the clusters that it forms, 
act together in the acceleration of the particles, whereby none of their two
power-law indices can be expected to be directly mapped onto the indices of the tail of the electrons' kinetic energy distributions, since the system is not linear and exhibits non-trivial complex dynamics, and there is also a dynamic effect, namely the particles' Lagrangian view-point, 
i.e.\ the electric fields a particle witnesses along its trajectory depend on the initial conditions. 

Analyzing the transport properties of the accelerated particles in energy space, we have shown that the Fokker-Planck transport equation is not consistent with the test particle dynamics inside the simulation box, for two reasons (1) the transport coefficients cannot be determined from the particle dynamics in a meaningful way, and (2) the random walk picture that is inherent in the FP approach does not apply. Of course, mathematically there may exist functional forms of transport coefficients that would reproduce the observed energy distribution when inserted into the FP equation, they would though be just formal and completely lack a consistent physical interpretation as diffusion and convection coefficients.  We have also shown that a fractional transport equation \citep{Isliker2017a, Isliker2017} is able to reproduce the acceleration process.
Its coefficients can consistently be determined from the particle dynamics, 
yet, despite its success in reproducing the particles' acceleration, there still is an inconsistency in its theoretical justification, since in its derivation it is also assumed that a random walk process (of the Levy type) takes place, see \cite{Isliker2017a}.
A completely adequate transport model for the case of single, yet power-law dominated and fractal acceleration, seems not to exist, to our knowledge.

The spatial transport of the heated plasma and the accelerated particles is anomalous, as one could expect it for the transport of particles in a fractal environment of electric fields. The high energy particles execute super diffusion in the initial phase of their acceleration and later become sub-diffusive. The opposite is true for the low energy, just heated particles, they start as sub-diffusive and end up in the final stage as super-diffusive. The squared spatial increments of the electrons increase with increasing squared energy increments, i.e.\ large spatial displacements are correlated with large energy gains. 
We must emphasize that 
	the combined and simultaneous 
	particle transport 
	(in position- and energy-space)
	inside a fractally distributed electric field environment, with the strength of the electric field moreover being power-law distributed, is a complex problem that requires further analysis 
	for an understanding of its nature.

The scenario emerging from the analysis presented here is related very nicely with current observations of energetic particles during the formation and eruption of jets: Initially, there is a slow evolution of the emerging flux till the point where the reconnecting large scale current sheet is formed \citep{Archontis13, Jiang16}. The spontaneous fragmentation of the current sheet creates a very efficient environment for the heating and acceleration of particles in the vicinity of the base of the standard and blow out jets. The high energy emission  recorded by the RHESSI satellite in the vicinity of jets confirms the presence of super hot plasmas, as we find it in our model.  Also, the simultaneous detection of type III and HXR bursts during impulsive explosions, and the correlations with Solar Energetic Particles, are related with jets 
\citep{Bain09, Glesener12, Glesener2018, Chen13, Raouafi16,Archontis2019}, 
and can directly be understood in the frame of our model as resulting from the acceleration on sub-second time-scales in the fragmented environments near jets, when embedded in the global topology of emerging flux, as in Fig.\ \ref{fl1} or \ref{fl1_53}.

The results reported here are based on the coupling of a resistive MHD code with a test particle code. This analysis has several advantages, under the prerequisite that the energy transfer from the electromagnetic fields to the heated and accelerated particles is limited to less than 15-20\% of the MHD energies. Also, no temporal interpolation is needed if the evolution of the MHD fields is very slow compared to the particle acceleration time. The analysis is indeed very useful since the statistical characteristics and the transport properties of particles can be analyzed on the kinetic level, and one can trace the evolution of the particle energy distribution in realistic, large-scale, open systems. The main disadvantage is that test-particle simulations cannot estimate the feedback of the accelerated particles on the evolution of the electromagnetic fields, therefore details of some of the  quantitative results (power law index of the high energy tail, maximum energy reached, acceleration time) of the saturated particle distributions must be expected to be revised to some degree when  feedback is included.  On the other hand, the basic statistical and transport properties studied here will persist and become a useful help and guide for the development of second generation  coupled MHD and PIC codes for large scale, open systems. These codes will be able to treat even more realistically particle heating and acceleration in the case of large scale explosions in the solar atmosphere. Several groups are working  towards this new generation of codes that couple  MHD and PIC simulations, but it is to early to evaluate their success \citep{Chen17,Drake18, Marle18, Makwana18}.

\acknowledgments
LV was partly supported  by the European Union (European Social Fund) and the Greek national funds through the Operational Program "Education and Lifelong Learning" of the  National Strategic Reference Frame Work Research Funding Program: Thales. Investing in Knowledge Society through the European Social Fund. 
VA acknowledges support by the Royal Society. 

\bibliography{MHD_Vasilis_2018}
\bibliographystyle{aasjournal}

\end{document}